\documentstyle[preprint,aps,epsf]{revtex}
\begin{document}
\draft
\title{Resonant spin current in nanotube double junctions}
\author{ Ryo Tamura}
\address{Department of Physics, Graduate School of Science, University
  of Tokyo, Hongo 7-3-1, Bunkyo-ku, Tokyo 113, Japan}
\maketitle

\begin{abstract}
 Zero bias conductance per spin of nanotube double junction (NTDJ) is investigated theoretically using the tight binding model, unrestricted Hartree-Fock approximation and non-equilibrium Green's functions.
 NTDJ consists of two metallic nanotubes joined by a piece of semiconducting nanotube, with the transition between the nanotubes made up of sets of 5 and 7 member carbon rings.
 A quantum well forms in the central semiconducting NT region, bounded by Schottky barriers.
 Spin current occurs when Coulomb interactions raise the spin degeneracy of resonant levels in the quantum well.
 As long as an appropriate semiconducting NT length is chosen, spin direction can be controlled by gate voltage, i.e., NTDJ functions as a nano spin filter.
\end{abstract}


The combination of nanotechnology based on carbon nanotubes (NTs) \cite{tube,tubeHamada} and spintronics \cite{spintronics} is thought to be extremely promising for future technological innovations.
 Tsukagoshi et al. have demonstrated efficient spin injection from cobalt electrodes in NTs.\cite{tsukagaoshi} 
 However, diffusion of magnetic atoms from the electrode into the NT results in an increase in uncontrollable spin flips due to the large spin-orbit interactions of ferromagnetic atoms.
Since long spin relaxation times are needed, spin filters composed of atoms with small spin-orbit interactions are desirable. 
In this Letter, a spin filter composed entirely of carbon atoms, arranged as a semiconducting NT between two metallic NTs is proposed.
 Schottky barriers at the interfaces form a quantum well within the semiconducting NT, while the metallic NTs function as leads.\cite{odintsov,tamuraMS}
 When Coulomb interactions lift spin degeneracy in the quantum well, resonant current becomes spin polarized.
 To the best of the Authors' knowledge, this is the first study into such
 resonant spin current.\cite{QWspin,QWnote}
 A possible method of constructing the NT junction is to allow open ends of $(3m,0)$ zigzag NT and $(3m \pm 1,0)$ NT to react; the NTs then become connected by adjacent regions of 5 and 7 member carbon rings, while still maintaining an sp$^2$ network.\cite{junctionSaito,junctionMeunier,junctionTamuraprb,junctionIijima}
 This structure is referred to as the NT double junction (NTDJ).
 In order to examine spin current in such devices, zero bias conductance per spin is calculated for (9,0)-(8,0)-(9,0) NTDJ.
 Results are closely related to the fourfold degeneracy of the central semiconducting NT.\cite{fourfold}

\begin{figure}

\caption{ 
 Atomic structure of nanotube double junction (NTDJ), showing an (8,0) NT connecting two infinitely long (9,0) NTs. 
Length of the (8,0) NT region, $L$ is $9a$, i.e., $L/(3a)=3$ where
 $3a$ is the length of the unit cell of (9,0) NT.
Note the 5 and 7 member carbon rings in the (8,0)/(9,0) transition region.
 The NTDJ is divided into regions $C$, $s$ and $d$, with regions $s$ and $d$ approximated as periodic.
  Gate surface and NTDJ's axis are taken to be parallel to each other, with a separation distance of five times the radius of (9,0) NT.
The length of (9,0) NT regions included in calculation region $C$ was 19.5$a$, although for illustrative purposes only 7.5$a$ is shown.
Geometries of the transition regions were taken to be the same, with the position where 5 and 7 member rings aligned assumed to be nearest to the gate.
}
\end{figure}

\begin{figure}

\caption{ 
Potential energy, i.e. diagonal terms of the converged Hamiltonian $H_c$, as a function of position across the left half of region $C$ for an NTDJ with $L=13$.
Values are averaged over each zigzag row.
}
\end{figure}

Using the unrestricted Hartree-Fock approximation, both the Hamiltonian $H^{\sigma}$ and the density matrix $\rho^{\sigma}$ are dependent on spin   $\sigma=\uparrow,\downarrow$, and can be expressed using the tight-binding approximation limited to the $\pi$ orbitals as \cite{harigaya}
\begin{equation}
H_{i,j}^{\sigma}=t_{i,j}-\tilde{U}_{i,j}\rho_{i,j}^{\sigma} 
+\delta_{i,j}\sum_{k,\sigma}\tilde{U}_{i,k}
(\rho_{k,k}^{\sigma} -(1/2)) \;\;,
\label{Hamil}
\end{equation}
where $ t_{i,j}= -t < 0$ when $i$ and $j$ are nearest neighbors,  $t_{i,j}=0$ otherwise.
NTDJ is divided into the two asymptotic 'lead' regions $s$, $d$, and a scattering region $C$, as shown in Figure 1, with Hamiltonian matrix elements assigned to one of these regions.
The lead regions are approximated as having a periodic structure based on the charges and Hamiltonian elements in the NT unit cells at the edges of region $C$, as indicated by rectangles in Fig.1.
 The charges contribute to diagonal terms of $H_c$, while Hamiltonian elements determine $H_{s,d}$ and $\tau_{s,d}$.
 For simplicity, only nearest neighbor elements are treated for off diagonal elements of $H_{s,d}$ and $\tau_{s,d}$.
 $s$ region, $d$ region and gate electrode voltages are referred to as $V_s$,  $V_d$ and $V_g$, respectively.
Since the gate electrode is a planar perfect conductor, and $V_g=0$ in our model, Coulomb interaction can be expressed as $ \tilde{U}_{i,j}=U(|\vec{r}_i-\vec{r}_j|) -U(|\vec{r}_i-\vec{R}_j|)\;,$ 
 where $\vec{R}_j$ is the image of $\vec{r}_j$.\cite{position}
 The 'bare' Coulomb kernel without a gate is assumed to be \cite{harigaya}
 \begin{equation}
 U(r)=2.5t/\sqrt{1+16(r/a)^2} \;\;.
 \label{U(r)}
\end{equation}
Voltages can be expressed in another way,
 $\tilde{V}_s=(V_s-V_d)/2$, $\tilde{V}_d=-\tilde{V}_s$
 and $\tilde{V}_g=-(V_s+V_d)/2$,
 since the difference between, not the absolute value of, voltages is of importance.
 In this letter, relative voltage representation is used and negative
 $\tilde{V}_g$ causing electrostatic hole doping is considered.

\begin{figure}

\caption{ 
   Conductance per spin as a function of gate voltage $\tilde{V}_g$ for $L=$13.
 $\epsilon_{\pm,\uparrow}$ and $\epsilon_{\pm,\downarrow}$ are shown by dashed and solid lines, respectively.}
\end{figure}

The density matrix $\rho$ is calculated from the retarded Green's function $G$, self energy $\Lambda^{\sigma}_p \equiv 
 \tau^{\sigma}_p(E+i\delta-H^{\sigma}_p)^{-1}\tau^{\sigma}_p $
 and its imaginary component $\Gamma^{\sigma}_p = i(\Lambda^{\sigma}_p-\Lambda^{\sigma*}_p)$, giving
\begin{eqnarray}
 G^{\sigma}= (E-H^{\sigma}_c-\Lambda^{\sigma}_{s}-\Lambda^{\sigma}_{d} )^{-1}\;\;,\nonumber \\
\rho^{\sigma}
=\frac{1}{\pi}\int^{\infty}_{-\infty}dE G^{\sigma}(\Gamma^{\sigma}_{s}
+\Gamma^{\sigma}_{d})G^{\sigma *}f_{\rm FD} \;\;,
\label{green}
\end{eqnarray}
 where  $ f_{\rm FD}= 1/(\exp(\beta(E-E_F))+1)$  and $k_B T = 1/\beta= 0.01t$.
\cite{dattabook}
 To calculate zero bias conductance, $E_F=eV_s=eV_d=|e|\tilde{V}_g$.
Using Eqs.(\ref{Hamil}) and (\ref{green}), $H$ and $\rho$ can be calculated  self-consistently for a given $\tilde{V}_g$.
 When self-consistency has been achieved, zero bias conductance for each spin $dI^{\sigma}/dV_{sd}$ is calculated from \cite{tomanek}
\begin{equation}
 \frac{dI^{\sigma}}{dV_{sd}}=  
\frac{\beta e^2}{h} \int_{-\infty}^{\infty} dE
 {\rm Tr}[\Gamma^{\sigma}_{s}G^{\sigma}\Gamma^{\sigma}_{d}G^{\sigma *}]
 f_{\rm FD}( 1- f_{\rm FD})\;\;.
\label{current}
  \end{equation}

 For small values of $|\tilde{V}_g|$, motion of holes into (8,0) NT is suppressed by the band-gap, resulting in more positively charged (9,0) NT
 than (8,0) NT and inducing Schottky barriers.
 Fig.2 shows potential energy as a function of position along the NTDJ for different spins and different values of $\tilde{V}_g$.
 It can be seen that as $\tilde{V}_g$ becomes more negative, the (8,0) NT region becomes electrostatically doped, lowering the Schottky barrier.
 Neighboring peaks and dips seen in the transition region between (9,0) NT and (8,0) NT are caused by an excess of electrons in the 5 member carbon rings, and a deficiency in the 7 member carbon rings \cite{junctionMeunier,Polytamura}.
 Note that while the gate electrode covers both $C$ and lead regions, and is much larger than a nanometer, it still exerts control over the atomistic spatial profile of the barrier.

Conductance per spin as a function of gate voltage $\tilde{V}_g$ is shown in Figs.3, 4 and 5 for $L=$13, 7 and 6 respectively, where $L$ is the length of (8,0) NT in units of $3a$.
 The first resonant conductance peak appears at the gate threshold, $\tilde{V}_{gth}$, with only a small tunneling current present when $\tilde{V}_g > \tilde{V}_{gth}$.
For smaller values of  $L$, enhanced tunneling effects increase hole density in the (8,0) NT region, making $\tilde{V}_{gth}$ more negative.
\begin{figure}

\caption{   Conductance per spin as a function of gate voltage $\tilde{V}_g$ for $L=$7.
 $\epsilon_{\pm,\uparrow}$ and $\epsilon_{\pm,\downarrow}$ are shown by dashed and solid lines, respectively.
When $\tilde{V}_g = -2.14 t$,  the spin polarization vanishes
 while the conductance peak appears.
 This peak is not due to $\epsilon_{\pm,\sigma}$
 but comes from  the level just under $\epsilon_-$ 
 denoted by $\epsilon'$. $\epsilon'$ is shown by the solid
 line with crosses only when $\tilde{V}_g < -2.14t$.}

\end{figure}
Fig.3 also shows the four eigenvalues of $H_c$ near the valence band edge,
 denoted by $\epsilon_{i,\sigma}$, where $i=\pm$ is $K$ corner point degeneracy.
 These can be represented as
 \begin{eqnarray}
\epsilon_{i,\sigma} &
= &
 \epsilon^{(0)}_{i}
+J_{(+,-)}n_{-i,-\sigma}  \nonumber \\
& & +(J_{(+,-)}-K)n_{-i,\sigma} 
+J_{(i,i)}n_{i,-\sigma}\;\;,
\end{eqnarray}
 where $J$, $K$, $n_{i,\sigma}$ and $\epsilon^{(0)}_{i,\sigma}$ are the Coulomb integral, exchange integral, occupation number and eigenvalue with no Coulomb interaction, respectively.
 Unlike finite systems or single electron tunneling,\cite{QWspin,Ohno}  $n_{i,\sigma}$ can take continuous values over the range 0 to 1.
 From Fig.3, it can be seen that large splitting between the levels of the different spin states only occurs when $\epsilon_{i,\sigma}$ crosses the Fermi level $E_F=|e|\tilde{V}_g$, i.e., when occupation numbers change.
 This indicates that splitting is a function of $J n_{i,\sigma}$ and 
 $K n_{i,\sigma}$, not of $\epsilon^{(0)}$.
Since the transmission rate $T(E)= {\rm Tr}[\Gamma^{\sigma}_{s}G^{\sigma}\Gamma^{\sigma}_{d}G^{\sigma *}]$
exhibits resonant peaks when $ E \simeq \epsilon_{i,\sigma}$, as $|\tilde{V}_g|$ increases two spin down peaks appear first, followed by two spin up peaks.
The pairing of peaks with common spin is due to Hund's rule, which holds when
  $K > \epsilon^{(0)}_+-\epsilon^{(0)}_{-}$.

 For a given $L$, barrier thickness $D$ decreases as hole number $N= 4-\sum_{i,\sigma} n_{i,\sigma}$ increases, as seen in Fig.2, broadening the energy of the resonant state, $W$.
 At the same time, $J_{(i,j)}$ decreases due to an increase in spatial width of the quantum well, $L-2D$.
 The net effect is an increase in $W - J_{(i,j)}$ resulting in a decrease in spin polarization.
 To examine spin polarization effects, total energy $E_{tot}$ is approximated by 
\begin{eqnarray}
E_{tot} &=&\sum_{i,\sigma}\left(\epsilon^{(0)}_{i}-\frac{W}{2}+\frac{W}{2}n_{i,\sigma} \right)n_{i,\sigma} + \sum_{i} J_{(i,i)}n_{i,\uparrow}n_{i,\downarrow} 
\nonumber \\ 
 &&-\sum_{\sigma} K n_{+,\sigma}n_{-,\sigma} + \sum_{\sigma,\sigma'} J_{(+,-)}n_{+,\sigma}n_{-,\sigma'} 
 \label{Etot}
 \end{eqnarray}
 Equation (\ref{Etot}) is derived from
 \begin{equation}
\Delta E_{tot}
 =  \Delta n_{i,\sigma}( W (n_{i,\sigma}-1/2)
+ \epsilon_{i,\sigma})\;\;,
 \end{equation}
 where the energy region $-W/2< E  -\epsilon_{i,\sigma} < +W (n_{i,\sigma}-1/2)$
 is occupied so that $n_{i,\sigma}$ changes at the energy $E=\epsilon_{i,\sigma}+W (n_{i,\sigma}-1/2)$.
 
 For $ 0 < N <1$, $n_{+,\uparrow}$ and $n_{+,\downarrow}$ initially decrease, because the degeneracy of $\epsilon^{(0)}$ is slightly lifted by 
 presence of
 pentagon-heptagon pairs making $\epsilon^{(0)}_{-}< \epsilon^{(0)}_{+}$.
 Therefore $n_{+,\uparrow}=1-\alpha$ ,$n_{+,\downarrow}=1-N+\alpha$ and $n_{-,\uparrow}=n_{-,\downarrow}=1$.
 Thus, $E_{tot} = (J_{(+,+)}-W)(N-\alpha)\alpha +C,$  with  $C$ independent of $\alpha$.
 For a given $N$, we define $\alpha_{min}$ as the value of $\alpha$ which minimizes $E_{tot}$.
 Thus, spin current occurs when $\alpha_{min}=0$.
 Although $\alpha_{min}=N$ is another solution, infinitesimal Zeeman energy favors $\alpha_{min}=0$.\cite{note}
 Thus the condition allowing spin down current at the first resonant peak $(N=1)$ is $W|_{N=1} < J_{(+,+)}|_{N=1}$.
 This condition is only satisfied when $L > 4$; since when $L$ is too small, large tunneling effects mean that $W-J$ is positive, even without electrostatic doping.
As $L$ increases, the value of $N$ over which polarization occurs also increases.
This can be understood by noticing that increases in $W - J_{(i,j)}$ per $N$ become smaller as $L$ becomes larger, because increases in hole density in (8,0) NT also decrease.
 
When $L>4$ and $1< N <2$,$( n_{-,\uparrow},n_{+,\uparrow};n_{-,\downarrow},n_{+,\downarrow}) = (1,1-\alpha,1-N_2+\alpha,0)$, where $N_2=N-1$ and $\epsilon_{-,\uparrow} < \epsilon_{-,\downarrow}$ on account of Hund's rule.
 Then
 \begin{eqnarray}
E_{tot} &= &(W-J_{(+,-)})\left(\alpha-\frac{N_2}{2}+
 \frac{\gamma}{W-J_{(+,-)}} \right)^2 +C \\
\gamma & = & ( \epsilon^{(0)}_{-}-\epsilon^{(0)}_+ +K +J_{(-,-)}
 -J_{(+,-)})/2 \;\;.
 \end{eqnarray}
 When  $\gamma $ is assumed to be positive and nearly constant, level configuration depends on the sign of $ W-J_{(+,-)}-(2\gamma/N_2) $.
 For a negative sign, holes are only injected into spin down levels, i.e.,  $\alpha_{min}=0$.
 Otherwise holes are injected into both spin levels, i.e., $\alpha_{min}=N_2/2-\gamma/(W-J_{(+,-)})$.
When $L=5$ or 6, $W-J_{(+,-)}-(2\gamma/N_2)$ increases, reversing sign as $N_2$ increases from 0 to 1.
 The increase in $W$ can be seen in the broadening of the peak corresponding to $(-,\downarrow)$ in Fig.5.
 As a result, spin up peaks overlap the second spin down peak, largely countering spin up polarization.
 For $L >6$, on the other hand, $W-J_{(+,-)}-(2\gamma/N_2)$ remains negative, and so the second spin down peak separates from the spin up peaks.

\begin{figure}
\caption{
 Conductance per spin as a function of gate voltage $\tilde{V}_g$ for $L=$6.
 $\epsilon_{\pm,\uparrow}$ and $\epsilon_{\pm,\downarrow}$ are shown by dashed and solid lines, respectively.
}
\end{figure}

 In addition to the lower limits of $L$ under which spin polarization can occur, upper limits also exist.
 As $L$ increases for fixed $N$, both thickness of the barrier $D$ and quantum well width $L-2D$ increase.
 The former reduces $W$ while the latter causes a decrease in the Coulomb integral $J_{(i,j)}$.
 When $J_{(i,j)}$ is less than either $W$ or $k_B T$, spin polarization disappears.
 Because $W$ decreases for larger $L$, $k_B T$ becomes a more important factor in determining the upper limit of $L$ than it was for the lower limit.
 Figure 3 shows that at room temperature, the upper limit is in excess of $39a$.

 In this letter, it has been shown that the nanotube double junction (NTDJ) functions as a nano spin filter without the use of magnetic atoms.
An infinitesimal magnetic field is necessary only to determine which spin becomes unoccupied first.
 Conditions for spin current are: (1) Fermi level is near the band gap edges at operating gate voltage $\tilde{V}_g$; (2) length of the semiconducting NT region, $L$, is larger than a particular threshold length.
 In fact, there are two threshold lengths, $L_1$ and $L_2$.
 For $L > L_2$, the direction of spin can be controlled by $\tilde{V}_g$.
 For $L_2 \geq L > L_1$, however, spin polarization is limited to a single direction.
 Furthermore there is no spin current when $ L_1 \geq L$.
 In the system studied, these thresholds are $L_1=4$ and $L_2=6$.
 Although the Coulomb interaction $U(r)$ of eq.(\ref{U(r)}) was chosen to be slightly smaller than estimated in Ref. \cite{harigaya} to achieve convergence of the self-consistent loop, the Author believes that the parameters used are realistic enough and results valid enough to prove that NTDJ is a useful device that combines nanotechnology and spintronics.

 The author wishes to thank Prof. Masaru Tsukada for valuable advice. 
 He also acknowledges Dr. Keiji Ohno for helpful discussions.
 This work was supported in part by a Grant-in-Aid for Creative Scientific
Research on "Devices on molecular and DNA levels" (No. 13GS0017) from the Japan
Society for the Promotion of Science.


\begin{thebibliography}{10}
\bibitem{QWnote}
 GaAs quantum well was treated as a spin filter
 in Ref.\cite{QWspin}, but only single charge tunneling
 with a fixed barrier was considered.
\bibitem{position}
The position of carbon atom $\vec{r}_j$ 
 can be determined approximately 
 from the shape of the original zigzag tubes.
This is justified since detailed atomic structure does not significantly modify electronic structure \cite{junctionMeunier}.
\bibitem{note}
 In the self-consistent calculation loop,
 the spin up Fermi level was chosen to be slightly higher
 than for down spin in the first iteration.
 For all other iterations, the same Fermi
 level was used for both spins.
\end{thebibliography}
\end{document}